\documentclass[epj]{svjour}

\usepackage{graphicx}
\usepackage{fancyhdr}
\def\makeheadbox{{
 }}
\def\mytitle{The NEMO~3 and SuperNEMO experiments} 
\def\myauthors{V. Vasiliev}    
\def\mytype{Contributed Talk}
\def\mysession{Flavor Physics}

\begin{document}
\title{Search for { $\rm \beta\beta$} decay with NEMO 3 and SuperNEMO experiments}
\author{V. Vasiliev 
\thanks{\emph{Email:} vv @ hep.ucl.ac.uk}%
 ~ on behalf of the NEMO~3 and SuperNEMO collaborations
}                     
%
%
\institute{University College London, Dept.
of Phys. and Astr., Gower Street, WC1E 6BT, London, UK
}
%
\date{}
\abstract{
NEMO 3 is a double beta decay experiment.
A part of low background data was analysed, preliminary result on
$\beta\beta_{2\nu}$ decay of $^{130}$Te obtained:\\
$T_{1/2}^{2\nu}(^{130}\rm Te) = 7.6 \pm 1.5(stat) \pm 0.8(syst) \times 10^{20} y$.
No $\beta\beta_{0\nu}$ signal was observed:
$T_{1/2}^{0\nu}( ^{100}\rm Mo) > 5.8 \times 10^{23}$ y and
 $T_{1/2}^{0\nu}( ^{82}\rm Se) > 2.1 \times 10^{23}$ y (90\% C.L.). SuperNEMO
 project R\&D has started.
} 
\maketitle
\section{Introduction}
\label{intro}

Neutrinoless double decay ($\beta\beta_{0\nu}$) is the most sensitive process for the search of
lepton number violation and its discovery would prove that the neutrino is a
massive Majorana particle. This process may occur through several mechanisms. In particular, the
existence of the $\beta\beta_{0\nu}$ decay by light neutrino exchange would allow to determination of the
mass scale of the neutrinos \cite{REVIEW}.

In the minimal supersymmetric standard model
 \\(MSSM) lepton number violation is forbidden by imposing the additional
discrete symmetry, R-parity. However, MSSM extensions with broken R-parity is another way
of lepton number violation in addition to the Majorana neutrino mass term. These models
lead to  $\beta\beta_{0\nu}$ decay via exchange of superparticles \cite{REVIEW}, squark
mixing \cite{Klapdor-SUSY} or neutrino mass generation via SUSY radiative corrections
\cite{Shimkovic-SUSY}.

The NEMO 3 is a double beta decay experiment running in the Fr\'ejus Underground
Laboratory in Modane, France. Its goal is to look for
neutrinoless double beta decay of $^{100}\rm Mo$ and $^{82}\rm Se$, as well as
to measure two neutrino double beta decay, $\beta\beta_{2\nu}$, for these and five other isotopes:
$^{116}\rm Cd$, $^{150}\rm Nd$, $^{96}\rm Zr$, $^{48}\rm Ca$ and
$^{130}\rm Te$. A facility to remove radon from the air around the detector was completed in October 2004.
About a year of low background data is being analysed now. A new preliminary result on $\beta\beta_{2\nu}$
decay of $^{130}$Te is reported.

An R\&D program is being carried out to produce a TDR for a new detector, SuperNEMO. It
will have a tracker, based on NEMO~3 design, and a calorimeter to fully reconstruct the events. 
It will have $\approx$100~kg of isotopes as a source. 
It is planned to reach a sensitivity of $\sim 1$--$2\times 10^{26}$ y for neutrinoless
mode of the decay.

\section{The NEMO 3 detector \cite{NIMNEMO}}

The NEMO~3 has a tracker to reconstruct the tracks of the two electrons from 
the $\beta\beta$ final state. It also has a calorimetery to measure their energy. 
This allows good discrimination between signal and background events.

The detector is cylindrical. A thin source foil ($\sim 50\; \rm mg/cm^2$) is
placed between two tracking volumes restricted by walls of plastic
scintillator blocks
read out by photomultipliers (PMT). In total about 10~kg of
$\beta\beta$ isotopes are installed in NEMO~3: 6.9~kg of $^{100}\rm Mo$, 0.93~kg of $^{82}\rm Se$,
 0.4~kg of $^{116}\rm Cd$,  0.45~kg of $^{130}\rm Te$,  37~g of $^{150}\rm Nd$,
 9~g of $^{96}\rm Zr$ and 7~g of $^{48}\rm Ca$.

The tracking volume consists of 6180 drift cells operating in Geiger mode. It provides
track vertex resolution of about 1 cm for 1 MeV electrons. The calorimeter has 1940 blocks
with an energy resolution at FWHM of $14$--$17\%/\sqrt{E}$. Its time resolution (250 ps) allows
excellent suppression of crossing electron
background with time of flight analysis.

There is a vertical magnetic field of 25 G to reject
$e^+e^-$ pair production in the
source. The whole detector is covered with 18 cm thick iron passive shielding and
30 cm thick tanks with boron water solution as neutron shield. The detector is capable 
of identifying $e^-$, $e^+$, $\gamma$ and $\alpha$ particles.

\subsection{Background model}

Background in NEMO 3 can be classified into three groups: (1) external from incoming $\gamma$,
(2) radon (Rn) in
the  tracker volume and (3) internal radioactive contamination of the source. All three were
estimated with event topologies other then $e^-e^-$ (i.e.  $e^-$, $e^-\alpha$, $e^-\gamma$, $e^-\gamma\gamma$ etc.)
using NEMO 3 data itself and verified with HPGe measurements and Rn detector measurements.

In the beginning of the NEMO 3 experiment it was found that the Rn level inside the tracking
gas is higher then initial requirements. It was identified that the main source of radon is the
laboratory air, which can penetrate the volume through small leaks. As a solution a facility was built
to purify the air surrounding the detector from Rn. This facility started to operate in October 2004.
Later we will refer to the data before that date as  "high background", or "Phase I" data. And to
"low background" or "Phase II" data for the period after October 2004.

\begin{figure}
\includegraphics[width=0.45\textwidth,height=0.45\textwidth,angle=0]{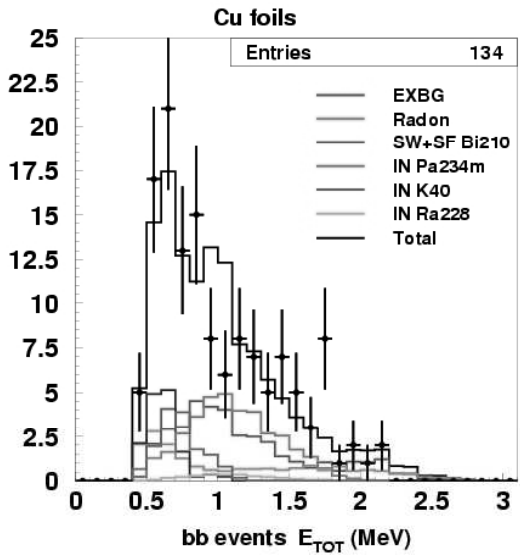}
\caption{\label{fgcu} Background two electron events
spectrum in Cu foil, Phase II data}
\end{figure}

To control the background, a blank Cu foil (621 g) was installed in the detector. In
Fig. \ref{fgcu} a spectrum
of two electron events measured in Cu as well as the MC prediction
according to the background model is
shown. It is an important check for our understanding of the background in the
experiment.

\section{NEMO 3 results}
\subsection {$\beta\beta_{2\nu}$ decay of  $^{130}$Te}

\begin{figure}
\includegraphics[width=0.45\textwidth,height=0.45\textwidth,angle=0]{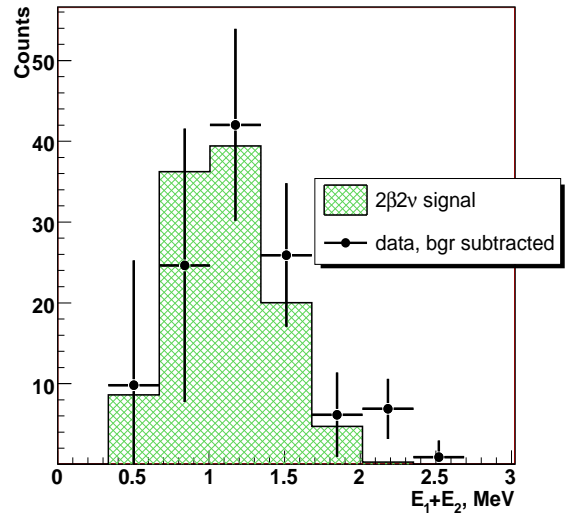}
\caption{\label{fgte} $^{130}$Te $\beta\beta_{2\nu}$ Phase II
  spectrum, background subtracted.}
\end{figure}

A preliminary measurement of $^{130}$Te half-life has been reported. Due to its extremely
low decay rate, only the low background data was used. In total, 534 days of data
was processed. The number of events observed from the Te source foil is 607, while
predicted background is 492 $\beta\beta$ events. A binned maximum likelihood method was used to analyse 
the data. A $\beta\beta_{2\nu}$ signal equal
to $109 \pm 21.5 (\rm stat)$ events was found, Fig.~\ref{fgte}. This corresponds to a
 half-life of
$T_{1/2}^{2\nu}(^{130}\rm Te) = 7.6 \pm 1.5(stat) \pm 0.8(syst) \times 10^{20}$~y.

This result is an improvement compared to the previous attempt of
$^{130}$Te decay direct measurement \cite{arnaboldi03}. It is also in agreement
with prediction based on the geochemical $^{82}$Se/$^{130}$Te ratio and present
 $^{82}$Se decay rate from direct experiments \cite{barabash}.

This result is of particular interest, because there is a disagreement between
geochemical measurements. One group of authors found $T_{1/2} \approx 8 \times 10^{20}$~y
\cite{geo_young}, while the other gives $T_{1/2}\approx 2.5--2.7 \times 10^{21}$~y \cite{geo_old}. Also
it was noticed that smaller $T_{1/2}$ value was obtained in the experiments with "young"
ores ($<$ 100 million years). This lead to the hypothesis that the differences can
be accounted for
variations of Fermi constant $\rm G_F$ with time
\cite{Gvariation}. NEMO 3 measurement
reported is not in contradiction of this hypothesis and suggests that possible $\rm G_F$ 
variation should be tested
using geochemical methods for other $\beta\beta$ decaying nuclei.

\subsection{$^{100}\rm Mo$ results}

\begin{figure}
\includegraphics[width=0.45\textwidth,height=0.45\textwidth,angle=0]{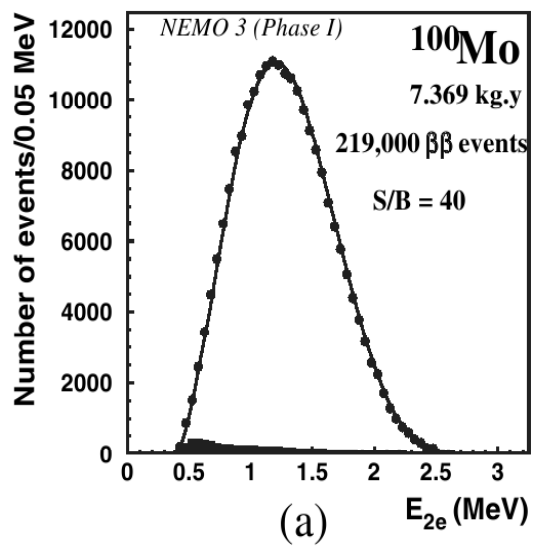}
\caption{\label{fgmo2n} $^{100}$Mo Phase I $\beta\beta_{2\nu}$ spectrum.}
\end{figure}

A measurement of $\beta\beta_{2\nu}$ decay was done with Phase I data, see
Table \ref{tb:2b2n} and Fig. \ref{fgmo2n}
\cite{NEMOPRL}. It is the biggest in the world number fo events was collected
(219000 events) with a signal to background ratio
(S/B) of 40. $T_{1/2}^{2\nu}(^{100}\rm Mo) = 7.11 \pm 0.02(stat) \pm 0.54(syst) \times 10^{18}$ y.

\begin{figure}
\includegraphics[width=0.45\textwidth,height=0.45\textwidth,angle=0]{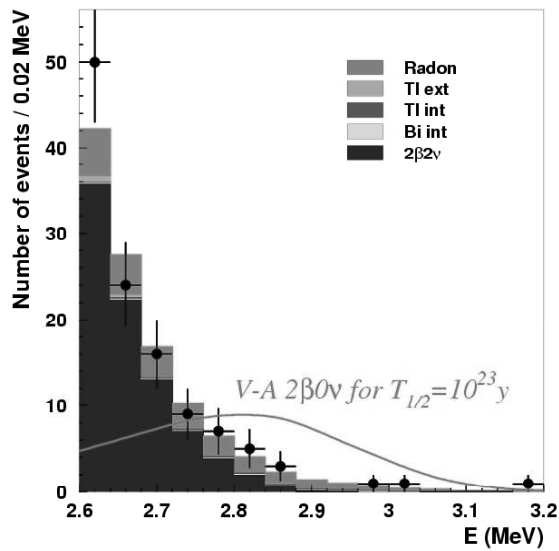}
\caption{\label{fgmo0n} $^{100}$Mo Phase I and Phase II spectrum at $Q_{\beta\beta}$.}
\end{figure}

No evidence for $\beta\beta_{0\nu}$ signal was found (Fig.
\ref{fgmo0n}).
The results for Phase I and Phase II data were combined. A preliminary
counting analysis shows 14 events in the window of interest [2.78--3.20]
MeV, the expected background is 13.4 events, and $\beta\beta_{0\nu}$
efficiency is 8.2\%. The effective time analysed is 13 kg$\cdot$y yielding a
lower limit on the half-life of $T_{1/2} > 5.8 \times 10^{23}$ y (90\% C.L.).

This corresponds to the effective Majorana neutrino mass $\langle
m_{\nu} \rangle < 0.8--1.3$ eV according to the most
recent QRPA NME calculations \cite{NME1}.

On the assumption of neutralino or gluino exchange, one can also derive the limit on supersymmetric trilinear coupling $\lambda_{111}^{'} < 1.5 \times 10^{-4}$
\cite{SUSYNME}.

\subsection{$^{82}\rm Se$ results}

2570 $\beta\beta_{2\nu}$ events from $^{82}$Se source were registered during Phase I of the
experiment, with a S/B ratio equal to  4, and a half-life given in Table \ref{tb:2b2n}.

After the preliminary analysis of 1.76 kg$\cdot$y of Phase I and Phase II data, 7
events were found in the window [2.62--3.20] MeV, with the expected background 6.4
events, $\beta\beta_{0\nu}$ efficiency of 14.4\% yielding a lower limit on the half-life of
$T_{1/2}>2.1\times 10^{23}$ y (90\% C.L.), which corresponds to an upper
mass limit of $\langle m_{\nu} \rangle < 1.4--2.2$ eV \cite{NME1}.

\subsection{$\beta\beta_{2\nu}$ decay}

Phase I data for four other isotopes were analysed and their half-lives
measured, see Table \ref{tb:2b2n} with preliminary
results reported. This is a very important input for
nuclear theory, because $\beta\beta_{2\nu}$ decay rate is used to fix some free parameters in QRPA nuclear models.
Since all isotopes are measured with the same device, the
half-life ratio has a very small systematic uncertainty, while statistical errors will
reach few per cent at the end of the experiment.

\begin{table}
\caption{ \label{tb:2b2n}
Main results on $\beta\beta_{2\nu}$ decays. S/B is a signal to
background ratio.}
\footnotesize\rm
\begin{tabular}{c c c}
\hline
Nuclei &  S/B & $T_{1/2}$, y\\
\hline
$^{100}\rm Mo$ & 40 & $[7.11 \pm 0.02 (\rm stat) \pm 0.54(\rm syst)] \times 10^{18}$ \\
$^{82}\rm Se$ & 4 & $[9.6 \pm 0.3 (\rm stat) \pm 1.0(\rm syst)] \times 10^{19}$ \\
$^{116}\rm Cd$ & 7.5 & $[
2.8 \pm 0.1 (\rm stat) \pm 0.3(\rm syst)] \times 10^{19}$ \\
$^{150}\rm Nd$ & 2.8 & $[
9.7 \pm 0.7 (\rm stat) \pm 1.0(\rm syst)] \times 10^{18}$ \\
$^{96}\rm Zr$ & 1 & $[
2.0 \pm 0.3 (\rm stat) \pm 0.2(\rm syst)] \times 10^{19}$ \\
$^{48}\rm Ca$ & $\sim$10 & $[
3.9 \pm 0.7 (\rm stat) \pm 0.6(\rm syst)] \times 10^{19}$ \\
$^{130}\rm Te$ & 0.25 & $[
7.6 \pm 1.5 (\rm stat) \pm 0.8(\rm syst)] \times 10^{20}$ \\
\hline
\end{tabular}
\end{table}

\subsection{Search for exotic processes}

Along with mass
mechanism, there are other possibilities to generate $\beta\beta_{0\nu}$ decay, e.g.
 if there is an explicit
 right-handed current (V+A) term in the Lagrangian, or if there are neutrino coupled
 axions (Majorons). In the first case one expects the
  angular and single electron energy distributions to
 differ from $\beta\beta_{0\nu}$ driven by
 the mass mechanism. Only a tracking detector like NEMO 3
 allows the use of this signature. In the second case axions are emitted in the decay, thus forming
 specific energy spectrum, characterised by spectral index $n$ \cite{NEMOMAJ}. One can look for
 deviations in the $\beta\beta_{2\nu}$ spectrum
 shape to restrict Majoron models and NEMO 3 is one of the best
 experiments for this work, taking into account purity and high number of 
 $\beta\beta_{2\nu}$ events collected.
In Table \ref{tb:exotic} results for (V+A) and Majoron search are summarised.

\begin{table}

\caption{\label{tb:exotic}
Constraints on $T_{1/2}$ in years for exotic
processes from NEMO 3 data (90\% C.L.). $\lambda$ is a (V+A) Lagrangian
parameter, $g$ is a Majoron to neutrino coupling strength; NME calculations
from \cite{NME1} were used. See \cite{NEMOMAJ} for explanation of
spectral index $n$.}
\footnotesize\rm
\begin{tabular}{c c c c c c}
\hline
Nuclei & $^{100}\rm Mo$ &  $^{82}\rm Se$ \\
\hline

(V+A) current & $>3.2\times {10^{23}} ^a$& $>1.2\times 10^{23}$&\\

 $n=1$&  $>2.7\times {10^{22}} ^b$& $>1.5\times 10^{22} $\\

 $n=2$&$>1.7\times 10^{22}$&$>6.0\times 10^{21}$\\
 $n=3$&$>1.0\times 10^{22}$&$>3.1\times 10^{21}$\\
 $n=7$&$>7.0\times 10^{19}$&$>5.0\times 10^{20}$\\
\hline
\multicolumn{3}{l}{$^a$ $\lambda < 1.8\times 10^{-6}$}\\
\multicolumn{3}{l}{$^b$ $g < (0.4-1.8)\times 10^{-4}$}\\
\end{tabular}
\end{table}

\section{SuperNEMO project}

The SuperNEMO collaboration has been formed in 2005
and started to study the feasibility of an extrapolation of the NEMO technique to a detector
with a mass of at least 100 kg of enriched $\beta\beta$ isotope. The goal is to reach a sensitivity of
50 meV on the effective Majorana neutrino mass.

SuperNEMO would use the NEMO~3 technical choices: a thin source between two tracking volumes
surrounded by a calorimeter. The performance characteristics to
improve, relative to the NEMO 3, are: the energy
resolution
(FWHM 7--10\%$/\sqrt{E}$ depending on final design is needed),
geometrical acceptance and $\beta\beta_{0\nu}$ detection efficiency,
the source radiopurity (factor $>10$ compared to NEMO 3)
and the background rejection techniques.

The collaboration focuses on two possible $\beta\beta$ sources: $^{82}$Se and $^{150}$Nd. $^{150}$Nd case has a
number of attractive features as 8 times bigger phase space factor and Q value above $^{222}$Rn daughters $\beta$-decay
\cite{REVIEW}. However,  it is difficult to enrich. Collaboration is studying the possibility to use MENPHIS facility in
France for $^{150}$Nd enrichment.

It is planned that the detector will have a modular structure. Each module containing 5--7 kg
of enriched isotopes, around 20 modules in total. If funded, the first module can start taking data as soon as
2010, with the whole detector finished by 2012--2013.

A  three
years R\&D program was approved in
the UK, France and Spain to achieve all these goals and make a detailed
technical design proposal by the end
of 2008.

%

\begin{thebibliography}{999}
%
%
\bibitem{REVIEW}
A. Faessler and F. \v Simkovic, J. Phys. G {\bf 24}, (1998) 2139;
J. Suhonen, O. Civitarese, Phys. Rep. {\bf 300}, (1998) 123.
\bibitem{Klapdor-SUSY}
M. Hirsch, H.V. Klapdor-Kleingrothaus and S.G. Kovalenko, Phys. Lett. B {\bf 372}, (1996) 181;
H. Pas, M. Hirsch and H.V. Klapdor-Kleingrothaus, Phys. Lett. B {\bf 459}, (1999) 450.
\bibitem{Shimkovic-SUSY}
M. Go\'zdz, W.A. Kaminski and F. \v Simkovic, Phys. Rev. D {\bf 70}, (2004) 095005.
\bibitem{NIMNEMO}
R. Arnold et al., Nucl. Instr. Meth. A {\bf 536}, (2005) 79.
\bibitem{arnaboldi03}
C. Arnaboldi et al., Phys. Lett. B {\bf 557}, (2003) 167.
\bibitem{barabash}
A.S. Barabash,  Czech. J. Phys. {\bf 56}, (2006) 437.
\bibitem{geo_young}
O.K. Manuel, J. Phys. G {\bf 17}, (1991) 221;
N. Takaoka, K. Ogata, Z. Naturforsch {\bf 21a}, (1966) 84;
N. Takaoka, Y. Motomura, K. Nagano, Phys. Rev. C {\bf 53}, (1996) 1557.
\bibitem{geo_old}
T. Kirsten et al., \textit{ Proc. Int. Symp. Nuclear Beta Decay and Neutrino (Osaka'86)} (World
Scientific, Singapore 1986) p.81;
T. Bernatowicz et al., Phys. Rev. C {\bf 47}, (1993) 806.
\bibitem{Gvariation}
A.S. Barabash, Eur. Phys. J. A {\bf 8}, (2000) 137.
\bibitem{NEMOPRL}
R. Arnold et al., Phys. Rev. Lett. {\bf95}, (2005) 182302.
\bibitem{ssd}
J. Abad et al., Ann. Fis. A {\bf80}, (1984) 9;
F. \v Simkovic et al., J. Phys. G {\bf27}, (2001) 2233.
\bibitem{NME1}
V.A. Rodin et al., Nucl. Phys. A {\bf 793}, (2007) 213;
M. Kortelainen and J. Suhonen, Phys. Rev. C {\bf 75},  (2007) 051303(R);
M. Kortelainen and J. Suhonen, Phys. Rev. C {\bf 76}, (2007) 024315;
M. Aunola and J. Suhonen, Nucl. Phys A {\bf 463}, (1998) 207.
\bibitem{SUSYNME}
A. Faessler et al., Phys. Rev. D {\bf 58}, (1998) 115004.
\bibitem{NEMOMAJ}
R. Arnold et al., Nucl. Phys. A {\bf 765}, (2006) 483.


\end{thebibliography}
%

\end{document}